\documentclass[aip,amsmath,amssymb,reprint]{revtex4-1}

\usepackage{graphicx}
\usepackage{dcolumn}
\usepackage{bm}

\usepackage[utf8]{inputenc}
\usepackage[T1]{fontenc}
\usepackage{mathptmx}
\usepackage{amsmath}  
\usepackage{amsfonts} 
\usepackage{epstopdf}
\usepackage{xcolor}
\draft 

\DeclareMathOperator*{\argmax}{arg\,max}
\DeclareMathOperator*{\argmin}{arg\,min}

\newcommand{\by}{$\times$}
\newcommand{\avg}[1]{\left\langle #1 \right\rangle}
\newcommand{\lrp}[1]{\left( #1 \right)}
\newcommand{\lrs}[1]{\left[ #1 \right]}

\begin{document}

\title{Precision measurement of tribocharging in acoustically levitated sub-millimeter grains} 

\author{Adam G. Kline, Melody X. Lim, Heinrich M. Jaeger}
\email[]{jaeger@uchicago.edu}
\affiliation{Department of Physics and James Franck Institute, The University of Chicago, 929 E 57th St., Chicago, Illinois 60637, USA}

\date{\today}

\begin{abstract}
Contact electrification of dielectric grains forms the basis for a myriad of physical phenomena. However, even the basic aspects of collisional charging between grains are still unclear. Here we develop a new experimental method, based on acoustic levitation, which allows us to controllably and repeatedly collide two sub-millimeter grains and measure the evolution of their electric charges. This is therefore the first tribocharging experiment to provide complete electric isolation for the grain-grain system from its surroundings. We use this method to measure collisional charging rates between pairs of grains for three different material combinations: polyethylene-polyethylene, polystyrene-polystyrene, and polystyrene-sulfonated polystyrene. The ability to directly and noninvasively collide particles of different constituent materials, chemical functionality, size, and shape opens the door to detailed studies of collisional charging in granular materials. 
\end{abstract}

\maketitle 

\section{Introduction}

In industry, electrostatic charging underpins manufacturing techniques such as powder coating~\cite{bailey1998science}, but can cause catastrophic explosions in the handling of fine powdered materials~\cite{abbasi2007dust}. 
The buildup of charge due to repeated collisions between small particles is thought to be responsible for lightning in volcanic ash clouds~\cite{brook1974lightning}, the electrification of grains in sand storms~\cite{stow1969dust}, and potentially also the very early stages of the formation of planetesimals from interstellar dust~\cite{blum2008growth,jungmann2018sticking}. Several controlling parameters for particle charging have been measured, notably the effects of particle size~\cite{forward2009charge, waitukaitis2014size}, atmospheric conditions and external electric fields~\cite{zhang2015electric,pahtz2010particle}, surface hydrophobicity~\cite{lee2018collisional}, frequency of contact~\cite{shinbrot2018multiple}, and kinetic energy prior to impact~\cite{poppe2000experiments}. However, the underlying mechanism for charge exchange remains an area of debate, particularly between insulators, which have very low charge mobility. 

Several candidates for the charge carrying species have been suggested, including electrons in trapped surface states~\cite{lowell1986triboelectrification, liu2008electrostatic, lacks2007effect}, ions in atomically thin water layers~\cite{mccarty2008electrostatic, lee2018collisional,harris2019temperature}, and mechanoradicals produced during contact~\cite{baytekin2011mosaic,baytekin2012really}. Elucidating the fundamental mechanism of collisional charging between grains thus calls for systematic, quantitative experiments. 
The most common experimental approach utilizes drop tests~\cite{mehrotra2007spontaneous,lee2015direct, waitukaitis2014size,jungmann2018sticking}, wherein a collection of grains is filmed and studied through the course of a free-fall, or in Faraday cup experiments~\cite{lamarche2009electrostatic, sowinski2009new} in which only net charges of collections of particles can be measured. These experiments typically involve a large number of particles, which results in both particle-wall and particle-particle collisions, hindering access to the basic physics of a single collision. Recent experimental techniques have made precise measurements of the impact charging of a single submillimeter particle with a fixed substrate~\cite{watanabe2006measurement,xie2016instrument,lee2018collisional,haeberle2018double}. There remains, however, a need to track the evolution of the charging process over repeated, highly controlled collisions between a pair of grains. 

Here, we introduce such a method by combining high-speed videography and acoustic levitation, allowing for the contact-free manipulation of a wide variety of constituent materials, particle sizes, and shapes~\cite{lee2018collisional,lim2019cluster,lim2019edges}. We dynamically control the location of stable levitation positions within the acoustic field, generating controlled collisions between a pair of particles. The charges on the particles can then be measured entirely non-invasively using the acoustic field, isolating the issues of granular charging to the repeated collisions between the particles. Our experimental protocol takes an important step towards capturing the dependence of charging on size, shape, material, spin, and collisional energy. 

Preliminary data demonstrates the utility of this approach with respect to tribocharging between a pair of grains of the same material. We show that pairs of polyethylene grains do not exchange significant charge over approximately one hundred collisions. In contrast, pairs of polystyrene particles exchange charge at a relatively constant rate of~$\approx 20 \: 000 \; e/$collision (in units of the elementary charge ~$e=1.6\times 10^{-19}$C). Sulfonating the surface of one of the polystyrene particles increased this charging rate by a factor of 10. 

Our results also suggest a general method to manipulate the location and number of potential minima in an acoustic trap. Previous designs for the transport of materials using acoustic levitation utilise highly structured acoustic interference patterns, through either highly coordinated inputs to a series of independently driven transducers~\cite{foresti2013acoustophoretic,courtney2014independent,marzo2015holographic,baresch2016observation}, or geometric patterns on the reflector and transducer surfaces~\cite{melde2016holograms,wang2016particle}. In contrast, our method requires only one transducer, with a single electrical drive, and the actuated motion of a boundary, to produce acoustic traps that can be reconfigured in real time.

\section{Experimental Setup}

The basic idea for the experiment is to actuate collisions between a pair of particles using ultrasound, separate them to their original positions, and then to extract the charge of the particles from their resonant oscillatory motion inside the trap. Thus the measurement sequence typically involves the following steps: 1) separately levitating two particles in the acoustic trap, 2) measuring their charge by applying an oscillating electric field, and 3) colliding the particles such that they return to their initial positions in the acoustic trap. After each collision, or after a series of such collisions, we measure the charge on each particle by repeating step (2). This sequence of events is repeated under computer control, allowing us to precisely track the charge transferred during the collision between a pair of grains. 

\subsection{Overview}

\begin{figure*}
\includegraphics[width = \textwidth]{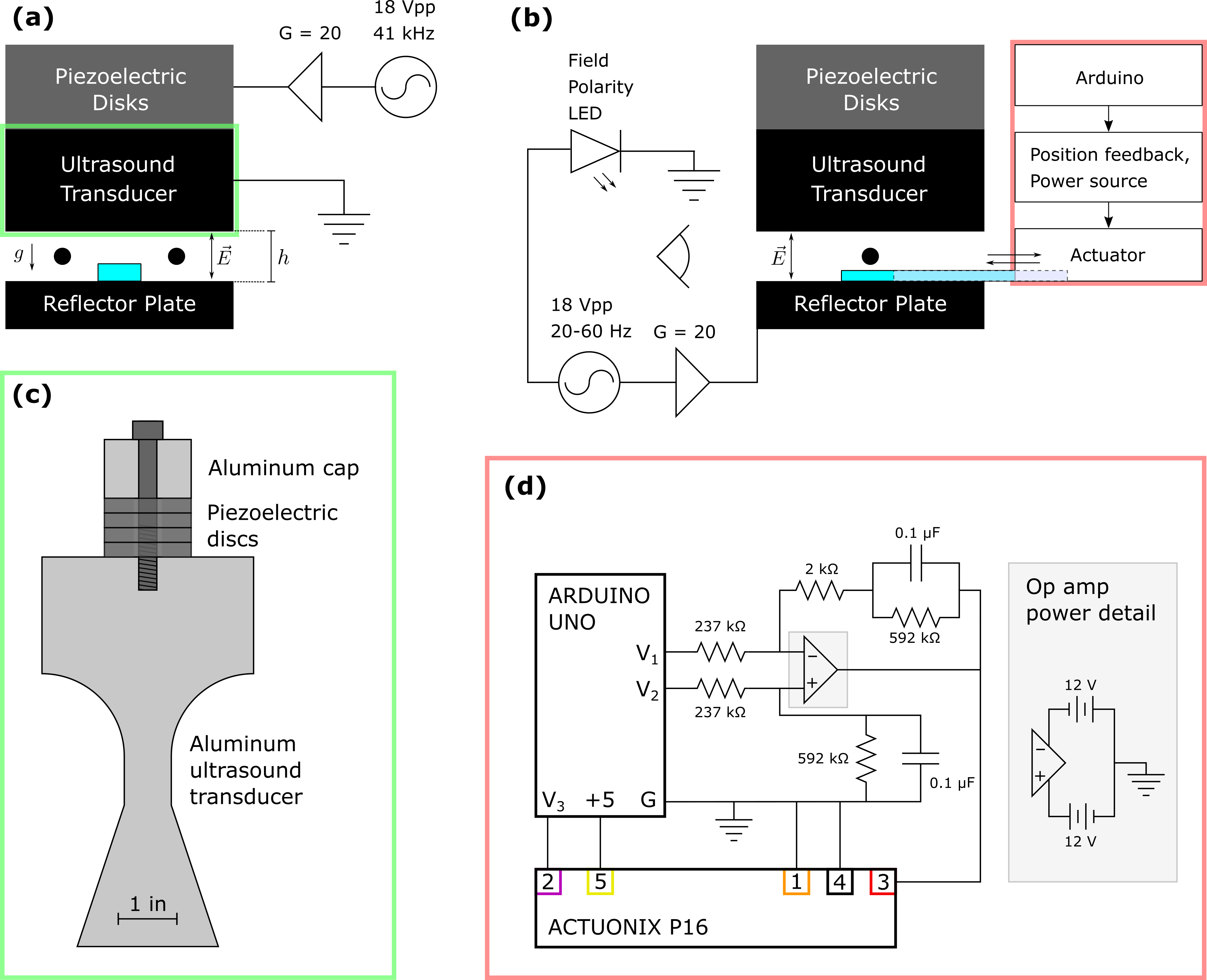}
\caption{ The experimental setup. (a) Schematic of the experiment from the perspective captured by the high-speed camera (front). A pair of submillimeter particles (black circles) are levitated halfway between a grounded aluminum transducer and an aluminum reflector. The reflector is mounted on a lab jack (Thorlabs), allowing precision control over the distance between transducer and reflector. Additionally, the reflector is connected to an AC voltage source, providing a vertical electric field across the gap between transducer and reflector. The particles are separated due to acoustic scattering from the acrylic ``hand'' (blue rectangle) between them, and are backlit and filmed from the front using a high-speed camera. The entire setup is enclosed in an acrylic chamber. (b) Scale-drawing of a cross-section of the (circularly symmetric) transducer. The transducer consists of four piezoelectric disks, bolted between an aluminium cap and the transducer base. The transducer base was designed using finite-element analysis software (COMSOL) to amplify the output of the disks, while also providing a spatially uniform driving amplitude at its base. (c) Schematic of the experiment from the side, showing the subsystems that control charge measurement (left) and particle collisions (right, red box). (d) Circuit diagram of the control system for the hand (and by extension, the collisions). An Arduino Uno drives a linear actuator by applying a~$\pm 5$V potential difference between two outputs, labelled~$V_1$ and~$V_2$. Position feedback from the actuator is then fed back into the Arduino (input~$V_3$), allowing for full control over the position of the hand.  
}
\label{fig:setup}
\end{figure*}

Figure~\ref{fig:setup}(a) displays a schematic of the experimental setup. A function generator (Agilent 3322a) and high-voltage amplifier (AA Labs A-301HS, gain 20) drive an ultrasound transducer at 41kHz in air (speed of sound~$c=343$m$/$s, wavelength~$\lambda\approx 8$mm) via piezoelectric disks (peak-to-peak voltage 360V). An aluminium plate, spaced a distance~$h$ beneath the transducer, acts as a reflector. Adjusting the spacing between the reflector and the base of the transducer to half the sound wavelength ($\lambda/2$) produces a standing wave with a single pressure node at~$\lambda/4$, in which particles can be levitated. We confine the particles to a one-dimensional track along the diameter of the transducer by machining a small channel ($l\times w\times d=$3.15\by50\by0.31 mm$^3$) in the reflector, and adjusting the distance between transducer and the top of the channel to match the resonance condition. 

A scale drawing of a cross-section of the aluminium transducer is shown in Fig.~\ref{fig:setup}(c). Four piezoelectric disks are bolted between an aluminium cap and transducer. Electrodes of alternating voltage are placed between the piezoelectric disks, such that both the cap and transducer base are grounded. These piezoelectric disks drive the base of the transducer. We designed the transducer shape using finite element analysis (COMSOL) to amplify the signal from the disks and produce a spatially uniform signal, resulting in a high-amplitude, roughly plane-wave ultrasound signal. The entire assembly has a resonant frequency of approximately 41kHz. The transducer and reflector are enclosed in an acrylic box with side-walls far from the levitation area (20"\by10"\by18") in order to mitigate the effect of side-wind perturbations. 

We measure the net charge on the levitated particles by applying an AC frequency-swept electric field between the aluminium reflector and (grounded) transducer. This vertical electric field is controlled by a second function generator (BK 4052) and high-voltage amplifier (AA Labs A-301HS, gain 20, shown schematically on the left side of Fig.~\ref{fig:setup}(b)), producing a total peak-to-peak voltage of 360V across the gap between transducer and reflector.  At the same time, we connect an LED in parallel with the (unamplified) output of the function generator. This LED produces a visible signal on the surface of the ultrasound transducer when the electric field is positive, allowing for direct visual access to the phase and frequency of the electric field throughout the experiment.  

In order to actuate collisions between particles in the acoustic trap, we move additional scattering surfaces within the acoustic field, thus dynamically changing the stable levitation locations within the trap. Specifically, we insert and withdraw a long, thin piece of acrylic (cross section 1.7\by7.6 mm$^2$, length 75 mm, shown in blue in Fig.~\ref{fig:setup}(a) and (b)) from the acoustic trap. When the acrylic ``hand'' is inside the trap, a pair of particles can be levitated on either side of the hand. In contrast, when the hand is removed, the particles each accelerate towards the centre of the trap, collide, and subsequently bounce. Reinserting the hand then separates the particles. This hand is attached to a linear actuator (Actuonix P16), which is in turn controlled by an Arduino Uno with a position feedback circuit (Figs.~\ref{fig:setup}(b) and (d)). The Arduino provides a positive (negative) 5 V difference on two pulse-width-modulated outputs to extend (withdraw) the hand from the trap. This signal is subsequently amplified and low-pass filtered, producing a $\pm 12$ V analog signal that drives the linear actuator. Position feedback from the actuator is then fed back to the Arduino, allowing for full control of the extension of the hand. 

The entire experiment (charge measurements, collisions, and data recording) is automated using Python, which actuates a collision by signaling the Arduino to extend and then retract the hand from the cavity. The timing of this process depends on the density of the levitated grains, since denser grains accelerate more slowly. For the particles used here (polyethylene and polystyrene), the actuator was set to retract 10mm at 32.5 mm/s, then immediately extend to its initial position at the same rate. Throughout the experiment, a high-speed camera (Phantom v12), also controlled using Python, records the motion of the particles (500 frames per second (fps) for the charge measurement, and 2000 fps for the collisions).  

\subsection{Controlled collisions}

\begin{figure}
\includegraphics[width = 0.4\textwidth]{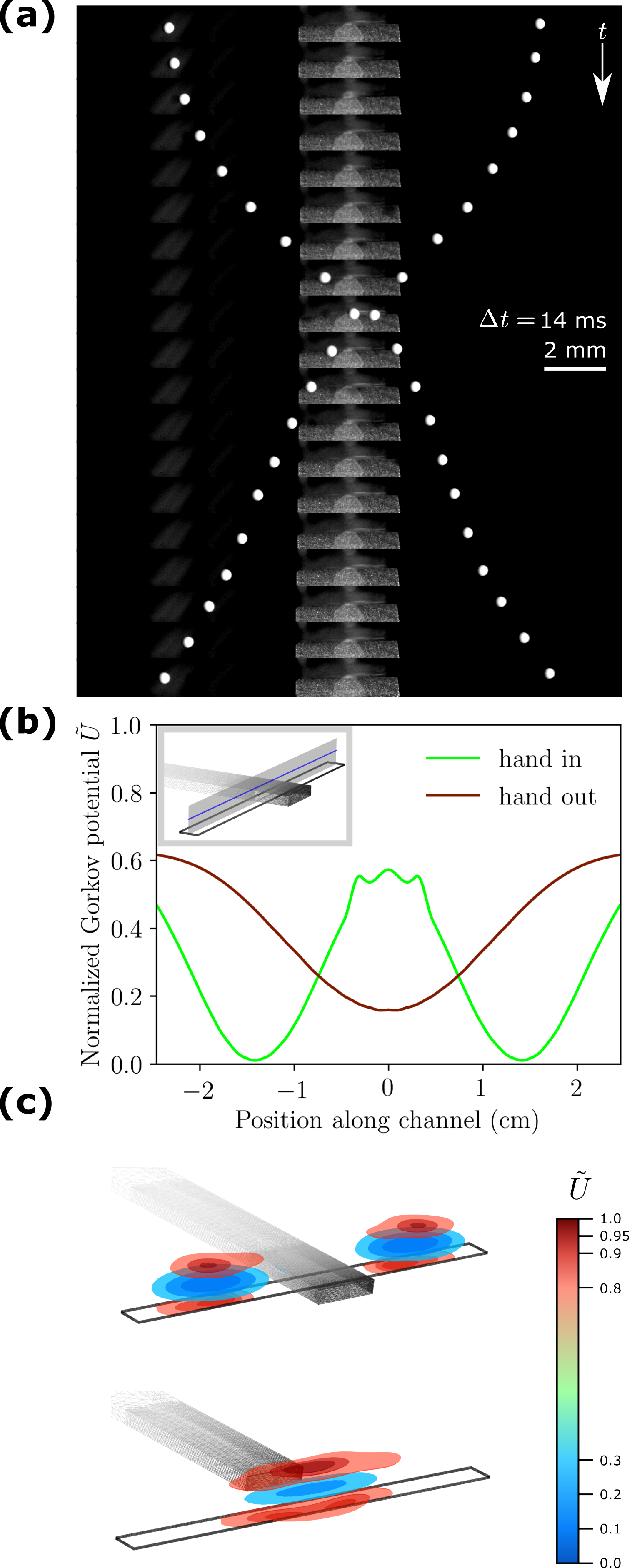}
\caption{Actuating binary collisions using the acoustic field. 
(a) Time-series of front-lit stills from the experiment, showing a collision between a pair of polyethylene particles (white). The grey rectangle in the center is the hand. 
(b) Gor'kov potential (scaled by the maximum acoustic potential, and offset such that the minimum acoustic potential corresponds to zero) experienced by a particle as a function of position along channel (blue line in inset). 
With the hand in, (green curve) two confining wells hold the particles separate, while removal causes these minima to coalesce in the center (red curve).
(c) Isosurface contours of the normalized Gor'kov potential for the two configurations: hand in (top) and hand out (bottom). The normalization for all data in (b) and (c) is the same.
Subfigures (b) and (c) were generated using finite element simulations (COMSOL).
}
\label{fig:collisions}
\end{figure}

Figure~\ref{fig:collisions}(a) shows a time-series of stills from the experiment, revealing the dynamics of a particle-particle collision. At the beginning of the collision, a pair of particles (white) levitate in the acoustic trap. The hand (centre gray rectangle) is retracted, causing the particles to accelerate towards the centre of the acoustic trap. They then collide and rebound. At the same time, the hand is reinserted into the trap, forcing the particles to return to their original positions on either side of the hand. 

In order to explain this result quantitatively, we consider the forces on the particles due to the presence and absence of the hand. Particles in the acoustic trap levitate and experience forces due to acoustic scattering. In the limit of particle radius~$R$ much smaller than the levitation wavelength ($R\ll \lambda$), this acoustic force is conservative, and can be expressed as the gradient of an acoustic potential. The shape of this potential is determined by the resonance of the cavity, which in turn depends on the geometry of the cavity and the location of scattering objects within it. 

Quantitatively, the acoustic potential can be calculated via a perturbation expansion of the acoustic fields due to scattering~\cite{Gorkov1961, bruus2012acoustofluidics7}, such that the acoustic potential~$U_\mathrm{rad}$ on a scatterer with radius~$R$, speed of sound~$c_p$, and material density~$\rho_p$ in an inviscid fluid with speed of sound~$c_0$ and density~$\rho_0$ is

\begin{align}
U&=\frac{4\pi}{3}R^3 \rho_0\left[f_1\frac{1}{2}c_0^2 \langle p^2\rangle - f_2\frac{3}{4}\langle v^2\rangle\right],
\label{eq:Urad}
\end{align}
where angled brackets denote time averages of the pressure~$p$ and velocity~$v$. The scattering coefficients~$f_1$ and~$f_2$ are given by
\begin{align*}
f_1&=1-\frac{c_p^2\rho_p}{c_0^2\rho_0}\\
f_2&=\frac{2(\rho_p/\rho_0-1)}{2\rho_p/\rho_0+1}.
\end{align*}

The pressure and velocity fields can thus be calculated within any trap geometry using finite element simulations, and then substituted into Eq.~\ref{eq:Urad} to predict the acoustic potential. We calculate the acoustic potential for our specific experimental trap geometry, with and without the hand, using COMSOL. In these three-dimensional simulations, we reproduce the experimental conditions (in the frequency domain) by driving the upper boundary with a normal displacement of 1 $\mu$m, then establishing perfectly reflecting boundary conditions on a parallel surface. We include the presence of a channel in the reflecting surface, with dimensions matching the experimental conditions (3.15\by50\by0.31 mm$^3$). The distance between the upper boundary and the reflector surface is set to $\lambda/2$. At the non-reflecting, lateral boundaries we impose plane wave radiation conditions.  We simulate the hand with a perfectly scattering rectangular block, located halfway along the channel (see inset of Fig.~\ref{fig:collisions}(b) for a diagram). 

Figure~\ref{fig:collisions}(b) illustrates the effect of the hand on the structure of the acoustic field. Plotting the acoustic potential as a function of the lateral position along the channel (blue line in inset of Fig.~\ref{fig:collisions}(b)) reveals that, when the hand is in the trap, the acoustic potential develops two distinct minima: a pair of particles can be levitated on either side. Alternatively, when the hand is removed from the trap, the two potential minima coalesce into a single minimum, located at the center of the trap. This change in the geometry of the acoustic potential forces particles to accelerate towards the center of the trap and collide. 

The shape of the acoustic potential in three dimensions confirms that particles can be stably levitated throughout the process of withdrawing and inserting the hand. Fig.~\ref{fig:collisions}(c) plots the isosurfaces of normalised acoustic potential~$\tilde{U}$ with (top) and without (bottom) the hand. In both cases, the acoustic potential wells are localized within the channel, and retain strong gradients in the vertical direction. Importantly, the acoustic potential wells are also well localized in the horizontal direction, ensuring that the particles remain stably trapped throughout the duration of the collision process. 

\subsection{Charge measurement}

\begin{figure*}
\includegraphics[width=\textwidth]{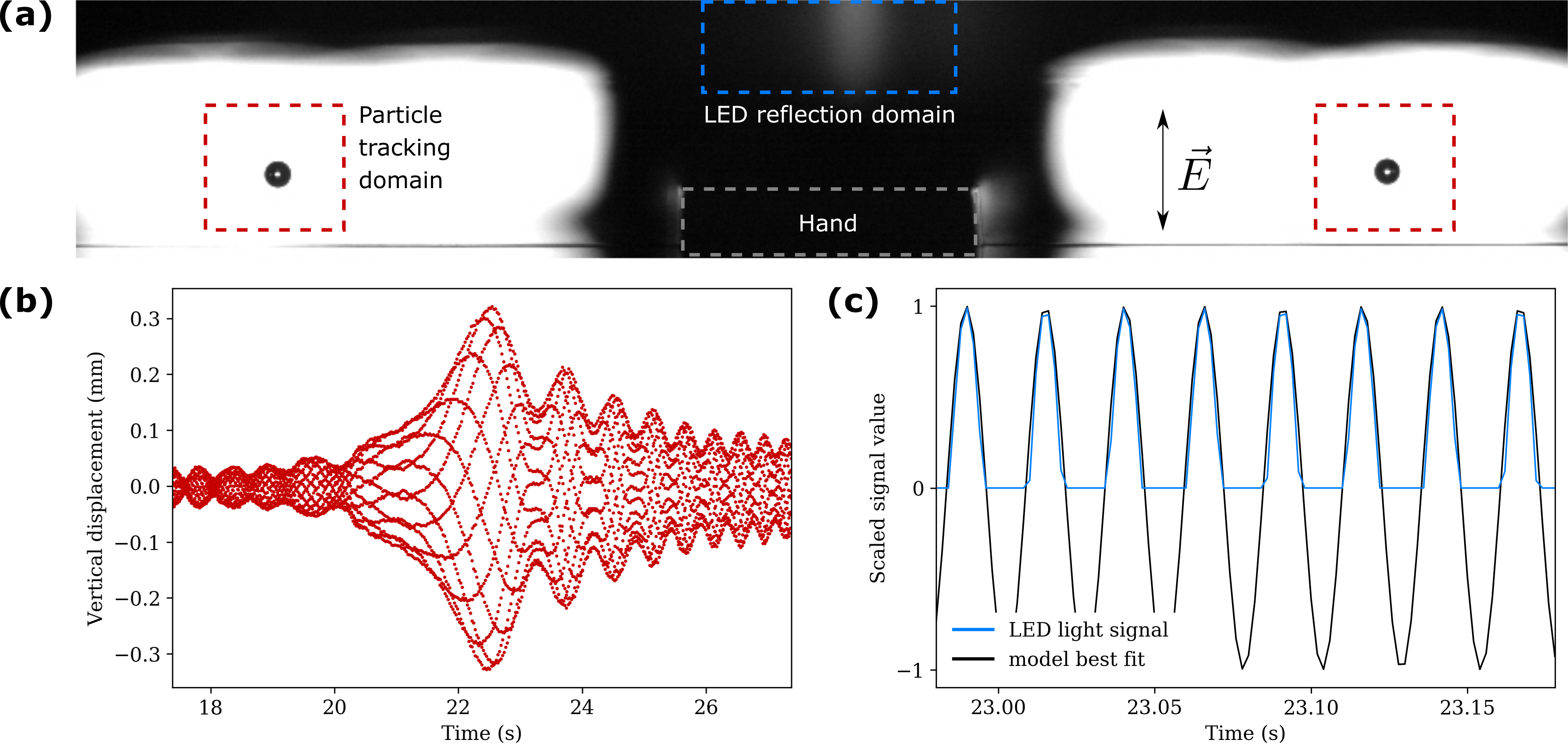}
\caption{ Measuring charges on individual particles using high-speed videography. 
(a) A still from the experiment, showing the (backlit) image captured by the high-speed camera during a charge measurement. A pair of particles (polystyrene, $D = $0.63 mm) are levitated on either side of the hand (outlined in gray box). The transducer is visible on the top of the image, with the light from the field polarity LED visible in the centre (bright patch). The dotted boxes indicate the separate parts of the image which are later analyzed: the trajectory of the two particles on either side (red boxes), and the intensity of the LED in the central box (blue dotted line). A bright LED corresponds to a positive (upward) electric field.  
(b) Experimental data for a section of the vertical particle displacement as a function of time,~$y(t)$, for a levitated particle in a frequency-swept AC electric field.~$y(t)$ here is measured from the median position of the particle, corresponding to its stable levitation position. 
(c) Experimental data for the brightness of the LED as seen on the surface of the transducer (blue line), normalised by its maximum brightness. The LED is connected in parallel with the electric field produced by the function generator, such that its brightness corresponds to the vertical component of the AC-swept electric field. The brightness of the LED is then fit to a frequency-swept sine wave (black line), recovering the phase and frequency of the electric field during the motion of the particles. 
}
\label{fig:charges}
\end{figure*}

In order to measure the charge, we apply a frequency-swept AC electric field~$\vec{E}(t)$ (sweep rate 0.5Hz/s)  across the gap between the transducer and reflector. This electric field oscillates each particle simultaneously. At the same time, the particles are subject to a vertical confining force due to the acoustic field (illustrated in Fig.~\ref{fig:collisions}(c)). The trajectory of each particle is thus a function of both its charge and the amplitude of the acoustic field. 

We ensure that the range of frequencies for the AC electric field includes the particle resonant frequency by manually setting the frequency limits for the first collision. During each subsequent collision, a Python script analyzes the oscillations of the particles during the previous charge measurement, and extracts the resonant frequency of the particles in the acoustic trap. The frequency range for the subsequent charge measurement is then adjusted such that the resonant frequency of a particle occurs two-thirds of the way through the frequency sweep. For the data shown in Fig.~\ref{fig:charging_rates}(a) (a pair of polyethylene grains), the size of the window was 15Hz; for the data shown in Fig.~\ref{fig:charging_rates}(b) and (c), the size of the window was 20Hz. 

 A back-lit still from the experiment during a charge measurement is shown in Fig.~\ref{fig:charges}(a). From the data, we extract the trajectories of the two particles (Fig.~\ref{fig:charges}(b)). We also measure the intensity of the LED connected to the electric field signal generator (Fig.~\ref{fig:charges}(c), data in blue). When~$\vec{E}(t)$ is positive, the LED shines with an intensity proportional to the amplitude of the electric field. Fitting this signal to the form of the applied frequency sweep (see Appendix B for details) thus allows for the instantaneous measurement of the phase and amplitude of~$\vec{E}(t)$ during the motion of the particles.
 
In order to determine the charge on the particle from its trajectory in the acoustic field, we start from the equation of motion for the vertical motion of the particle, with mass~$m$ and charge~$q$: 
 \begin{align}
     m\ddot{y}&=-mg -F_d+F_{ay}+qE_0\sin{(\omega_E(t)t)}\,.
     \label{eq:EOM}
 \end{align}
Here~$mg$ is the force due to gravity,~$F_{d}$ the air drag~$F_{ay}$ the acoustic force in the vertical direction, and~$E_0$ and~$\omega_E(t)$ are the amplitude and frequency of the applied frequency-swept electric field. 

To derive~$F_{ay}$ in Eq.~\ref{eq:Urad}, we consider the acoustic velocity potential for a standing wave in the~$y$-direction, with the reflector plate set to~$y=0$:

\begin{align}
    \phi(y,t)&=-\frac{v_0}{k}\cos{ky}\sin{\omega t}\,,
\end{align}
where~$v_0$ is the maximum acoustic velocity,~$k=2\pi/\lambda$ is the wavenumber, and~$\omega=kc$ is the angular frequency of the sound. The acoustic pressure and velocity in the~$y$-direction are thus

\begin{align}
    p&=-\rho \frac{\partial \phi}{\partial t}=\rho c v_0 \cos{ky}\cos{\omega t}
    \label{eq:pvertical}
\end{align}
and 
\begin{align}
    v&= \frac{\partial \phi}{\partial y} = v_0 \sin{ky}\sin{\omega t}.
    \label{eq:vvertical}
\end{align}

Substituting Eqs.~\ref{eq:pvertical} and~\ref{eq:vvertical} into Eq.~\ref{eq:Urad}, then taking the derivative, yields an expression for~$F_{ay}$: 

\begin{align}
    F_{ay}=\frac{5}{8} mv_0^2 k\sin{2ky}\,.
    \label{eq:Fay}
\end{align}

COMSOL simulations confirm that this expression for ~$F_{ay}$ is accurate even in the presence of the hand and channel. 

We model the force due to air resistance with the form

\begin{align}
    F_d &= 2m\beta_0 \dot{y} + 2m\beta_1 |\dot{y}|\dot{y}\,,
    \label{eq:Fd}
\end{align}
where the coefficients~$\beta_0$ and~$\beta_1$ are fitting parameters to be derived from the measured particle trajectory. In total, there are four fitting parameters to be derived from the particle trajectory: the particle charge $q$, the acoustic amplitude $a \equiv \frac{5}{8}v^2_0 k$, and the air drag coefficients $\beta_0$ and $\beta_1$. 

The full equation of motion for the charged, acoustically levitated particle in a frequency-swept AC electric field is then

 \begin{align}
     \ddot{y}&=-g + a \sin{2ky} - 2\beta_0 \dot{y} - 2\beta_1 |\dot{y}|\dot{y}+\frac{qE_0}{m}\sin{\omega_E(t)t}\,.
     \label{eq:EOMfull}
 \end{align}
 Given the trajectory~$y(t)$, its derivatives~$\dot{y}$ and~$\ddot{y}$, and the instantaneous phase and amplitude of the electric field, Eq.~\ref{eq:EOMfull} is linear in the four unknown constants~$a$,~$\beta_0$,~$\beta_1$, and~$q$. We thus measure the charge by performing a linear regression on the complete trajectory data from the experiment, and extracting the four unknown parameters from the coefficients of the fit. 
 
\section{Charge fitting procedure}
Our basic strategy is to treat the acceleration terms on the right hand side (RHS)--those that are functions of position, velocity, and time--as independent variables, and treat the total acceleration as the dependent variable. To see how this can be solved with regression, we begin by re-writing Eq. \eqref{eq:EOMfull} in vector representation. We write the set of unknown physical parameters in the form $\vec{\theta} = (a,\beta_0,\beta_1,q)^T$. For a given estimate of $\vec{\theta}$, the equation of motion can be written at each time point $t_j$ as
\begin{align}\label{eq:EOM_single}
    Z_j &= \ddot{y}(t_j) + g = \vec{X}(t_j) \cdot \vec{\theta} + A(t_j)\epsilon(t_j)
\end{align}

where
\begin{align*}
    \vec{X}(t_j) \cdot \vec{\theta} &= aX_1(t_j) + \beta_0 X_2(t_j) + \beta_1 X_3(t_j) + q X_4(t_j)\,.
\end{align*}

Here we have moved the acceleration due to gravity~$g$ to the left hand side (LHS) because it is known \textit{a priori}. Additionally, because~$\vec{\theta}$ is time independent, we must include a time dependent error~$\epsilon(t_j)$. According to this construction, each~$X_i(t_j)$ is one term on the RHS of Eq. \eqref{eq:EOMfull} corresponding to its coupling~$\theta_i$. For example,~$X_1$ corresponds to~$\theta_1 = a$, so~$X_1(t_j) = \sin 2ky(t_j)$. By including the error scaling in a diagonal matrix~$\Omega_{jj} \equiv 1/A(t_j)$, we can rewrite Eq. \eqref{eq:EOM_single} for all~$t_j$ simultaneously as 
\begin{align}
    \vec{Z} &= X^T\vec{\theta} + \Omega^{-1}\vec{\epsilon}\,.
    \label{eq:EOM_vectorized}
\end{align}

The error scaling $\Omega$ corrects for the fact that the error variance in the un-scaled case is proportional to the squared amplitude of the trajectory~$A(t_j)^2$ (heteroscedasticity). As a result, we statistically weight the data to increase the significance of those data where the oscillation amplitude is small (lower noise)\cite{fox1997applied}. The best linear unbiased estimate for $\vec{\theta}$ can then be found by minimizing $\epsilon^2$, the sum of squared residuals, or equivalently by enforcing $X\Omega\vec{\epsilon} = 0$. This yields
\begin{align}
    \vec{\theta} &= (X\Omega^2 X^T)^{-1}X \Omega^2 \vec{Z}\, .
    \label{eq:regression}
\end{align}

A successful measurement of~$\vec{\theta}$ hinges on a maximally accurate measurement of all elements of~$X$ and~$Z$. We outline three corrections to common issues with the data here. First, the acoustic acceleration term of Eq.~\eqref{eq:EOMfull} depends sensitively on the absolute height of the particle above the reflector, which cannot be straightforwardly extracted from the video. In order to correct for this unknown offset to the trajectory data, we perform linear regression as in Eq.~\eqref{eq:regression}, but where the position data is shifted by some amount~$\varphi$, for some small range of~$\varphi$. We then take the~$\varphi$ that minimizes the residuals of the regression to be an estimate for the true height offset. For a full discussion of the technical details, including the range over which~$\varphi$ was varied, see Appendix A. 

Second, measuring the acceleration of the particle due to the electric field depends on an exact measurement of the instantaneous electric field. Since the charge~$q$ is only coupled to the trajectory through the electric field, any measurement error in the electric field leads to a loss of precision in the charge measurement. In the current experiment, there is some error due to a lack of synchronization between the camera and the function generator. We account for this error by fitting the signal from the LED to the input signal from the function generator, with a phase delay~$\psi$ and apparent timescale~$t_s$ (details in Appendix B). This allows us to circumvent the lack of synchronisation between the camera clock and the function generator clock, thus measuring the true instantaneous electric field. Future experiments could avoid this issue by synchronizing the two clocks, or by synchronizing an oscilloscope with the camera clock and directly measuring the driving voltage. 

Third, measuring the damping terms and the LHS of Eq.~\eqref{eq:EOMfull} requires taking derivatives of the position data. The trajectory of each particle is approximately harmonic, with a similar frequency to the frame rate of the camera (the camera frame rate is less than a factor of 10 greater), such that finite difference estimates for the velocity and acceleration underestimate the true values. We correct for this by re-scaling the estimated velocity and acceleration (see Appendix C for details). Filming at a higher frame-rate would reduce the effect of this error.   
    



\begin{figure}
    \centering
    \includegraphics[width = 0.48\textwidth]{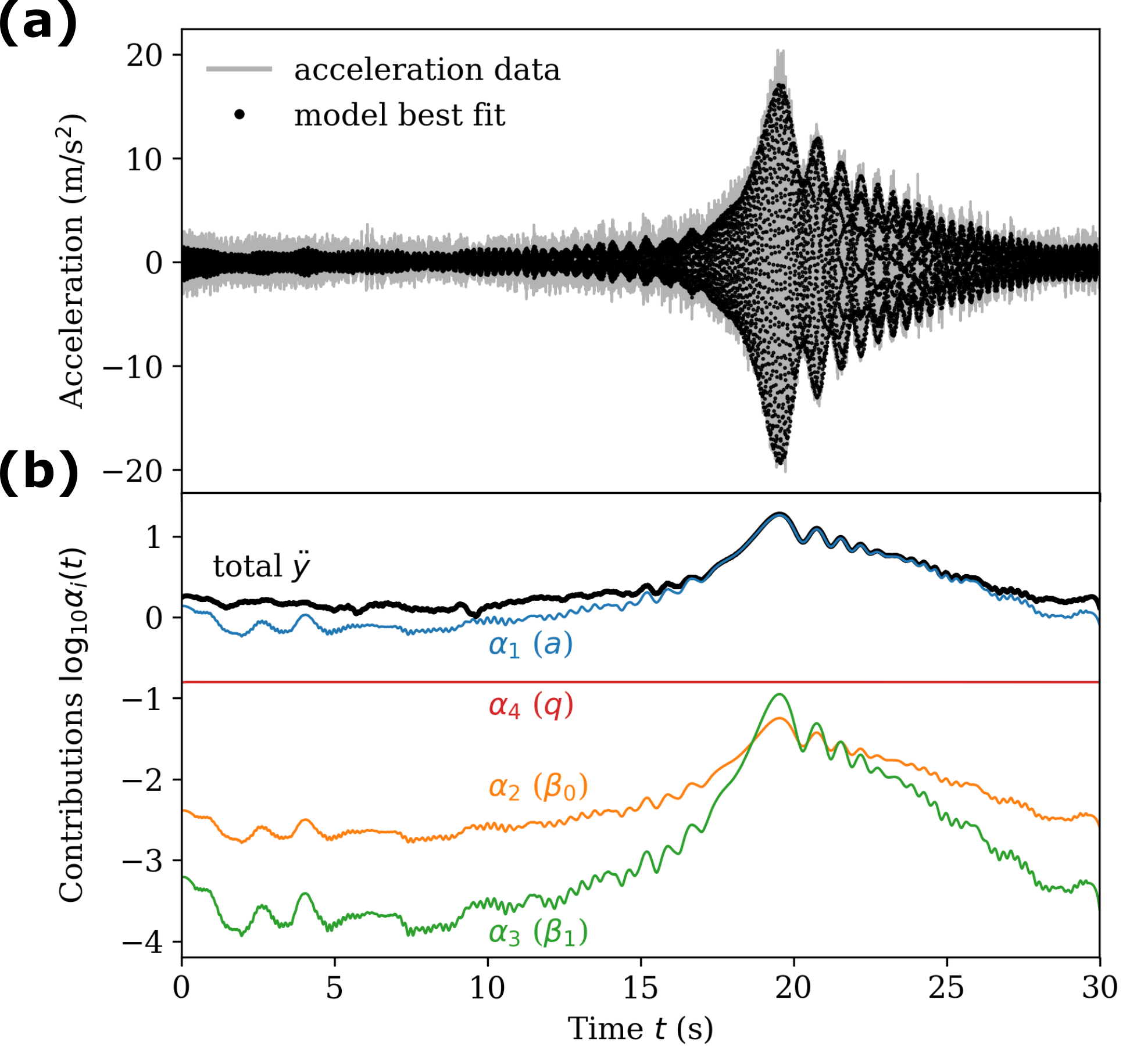}
    \caption{Demonstration of equation of motion fitting using Eq.~\eqref{eq:regression}. (a) Measured acceleration data (gray line) are compared to the best fit equation of motion (black points). (b) Amplitude of the contribution of each term on the right hand side of Eq.~\eqref{eq:EOMfull} to the total acceleration (black line) as a function of time. We plot the acoustic acceleration (blue,~$\alpha_1$), electric field acceleration (red,~$\alpha_4$), linear velocity-dependent damping (orange,~$\alpha_2$), and nonlinear damping (green,~$\alpha_3$). The amplitudes of the signals were extracted using a Hilbert transform. 
    } 
    \label{fig:charge_fitting}
\end{figure}

 Figure~\ref{fig:charge_fitting}(a) compares the (corrected) experimental acceleration data,~$\vec{Z}$ (gray lines), with the best fit to the model (Eq.~\eqref{eq:regression}),~$X^T\vec{\theta}$ (black points). We find generally good agreement over the entire time-series, with the largest errors appearing for large positive accelerations (near~$t=20$ in Fig.~\ref{fig:charge_fitting}(a)). This excess error appears in general for high-amplitude trajectories, suggesting that the forces on the particle near the channel are stronger than predicted by our ansatz, Eq.~\eqref{eq:Fay}. 

In order to ascertain the impact of this excess error, we plot the contributions of the four fitted forces to the totally acceleration data (Fig.~\ref{fig:charge_fitting}(b)). Throughout the trajectory, the acoustic force ($\alpha_1$, plotted as a blue line) contributes most significantly to the final result: the acoustic force is measured with the highest certainty for all parts of the reconstructed trajectory. The damping forces ($\alpha_2$, plotted in orange, and $\alpha_3$, green) contribute most strongly when the amplitude of oscillation is large. In contrast, the acceleration due to the electric field ($\alpha_4$, plotted in red) is constant throughout the sweep, and can therefore be measured most accurately in the low amplitude parts of the trajectory, when the contributions from the other forces are proportionally smaller. The deviation of the fitted trajectory from the data at high amplitudes thus has only a relatively small effect on the measured charge. The effect of this excess error on the charge measurement is further reduced by the heteroscedastic weighting of the data points, which statistically favors points with small acceleration amplitude.  


\section{Example data}

\begin{figure*}
\includegraphics[width=\textwidth]{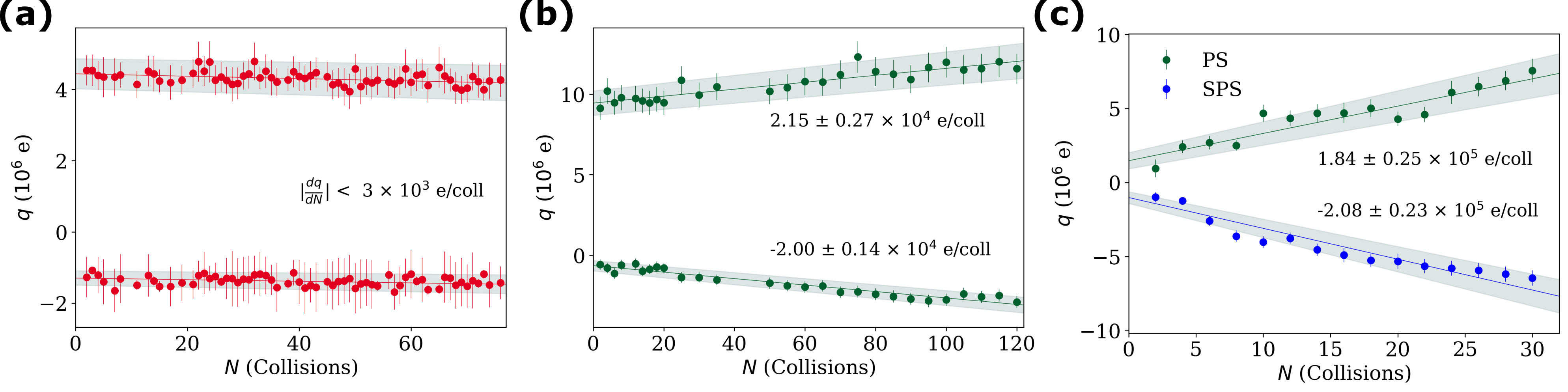}
\caption{Three examples of charging data obtained using our method, with error estimates for charges and charging rate given at 1 standard deviation, as calculated from Eq.~\eqref{eq:err_q}. (a) Data from 76 collisions between two polyethylene (PE) grains, showing no significant charging within experimental uncertainty. (b) Data from 120 collisions between two polystyrene (PS) grains. (3) Data from 30 collisions between two different grains, one polystyrene and one sulfonated polystyrene (SPS).}
\label{fig:charging_rates}
\end{figure*}

As a demonstration of the generality of the charge measurement procedure, we collided several types of particle. We began with commercially available polyethylene particles (Cospheric, material density 1 000 kg m$^{-3}$, diameter 710-850$\mu$m), and polystyrene particles (Norstone, material density 1 050 kg m$^{-3}$, diameter 620-780$\mu$m). In addition to the bare particles, we also sulfonated the polystyrene particles following the procedure described in Ref.~\cite{coughlin2013sulfonation}: 4g polystyrene was added to a vessel with 40 mL pure sulfuric acid. This mixture was stirred at 60~$^\circ$C for one hour, then removed from heat and rinsed thoroughly with DI water. 

Charging time-series for three types of particle-particle collision are shown in Fig.~\ref{fig:charging_rates}. In order to estimate the error associated with the charge fitting, we consider the uncertainty in the charge for a single charge measurement (measuring and fitting the response of a particle to a single AC frequency sweep). From standard weighted least squares regression, the uncertainty~$\Sigma_\theta$ in the fitting parameters~$\vec{\theta}$ is given by 

\begin{align}
    \Sigma_\theta &\approx \Sigma_s + (X\Omega^2X^T)^{-1}\frac{\epsilon^2}{N_t}
    \label{eq:regression_err} \, .
\end{align}

This expression includes the uncertainty in the conversion from the measured size of an object to its actual size (pixel scale error~$\Sigma_s$, see Appendix D for more details), as well as the measurement error in the trajectory itself ($\epsilon$, with total number of data points~$N_t$), propagated through the regression. On the basis of this expression, the uncertainty in the measured charge for a single data point is thus the square root of the matrix element of~$\Sigma_\theta$ associated with~$q$:

\begin{align}\label{eq:err_q}
    \delta q = \sqrt{(\Sigma_\theta)_{44}} \, .
\end{align}

Equation~\ref{eq:err_q} provides the error on an individual charge measurement after a collision. These errors are plotted as vertical lines around each point in Fig.~\ref{fig:charging_rates}. From the charge time-series, we can also extract the rate of charge transfer between a pair of grains,~$dq/dN$. In general, we find that each particle charges by a constant amount with each collision. We thus fit the charging time-series to a line using linear regression (plotted as solid lines in Fig.~\ref{fig:charging_rates}), such that the slope of the line gives~$dq/dN$. The standard error of the fit, which combines the scatter in the data with the error on the individual data points, is plotted as a gray shaded area. See Appendix A for a detailed derivation of the standard error of the charging rate. 

Within the error of the data, we find that polyethylene particles do not exchange significant charge over the scale of 76 collisions (Fig.~\ref{fig:charging_rates}(a)), with measured charging rate~$dq/dN$ smaller than $3000 e/$collision. In contrast, colliding a pair of polystyrene particles (Fig.~\ref{fig:charging_rates}(b)) produces a charging rate of $dq/dN\approx 20 \: 000\: e$/collision. This charging rate is highly consistent: even when the particles were allowed to collide several times in between charge measurements, the charge time-series follows the same linear trend. Sulfonating one of the polystyrene particles (Fig.~\ref{fig:charging_rates}(c)) enhanced the rate at which particles exchanged charge by a factor of almost 10, with measured charging rate of $dq/dN\approx 200 \: 000\: e$/collision, suggesting a link between surface chemistry and the propensity to exchange charge. 

Throughout each collision-series, the total charge of the particles is conserved within experimental error, in line with previous findings~\cite{lee2018collisional} that levitated particles exchange charge only during collisions (the particles do not exchange charge with the ambient gas).  In addition, the lack of charge saturation implies that our particles contact each other at slightly different spots each time, in agreement with previously reported trends~\cite{lee2018collisional,harris2019temperature}. Based on these two observations, we can infer an average surface charge density. By measuring the collisional velocity of the particles during the experiment, we estimate a maximum contact area~$A_c\sim 1000 \mu\mathrm{m}^2$ (see Appendix B for details). This corresponds to an average transferred surface charge density of~$3e/\mu\mathrm{m}^2$,~$20 e/\mu\mathrm{m}^2$, and~$200e/\mu\mathrm{m}^2$ respectively for the three data sets shown in Fig.~\ref{fig:charging_rates}, which are comparable to previous studies of granular tribocharging~\cite{mccarty2008electrostatic,haeberle2018double}.

\section{Conclusions}

We have constructed an experimental method capable of noninvasively triggering repeated controlled collisions between grains, and then measuring their individual electric charge with high precision. Our experiment is the first demonstration of a tribocharging experiment where the grains are completely isolated from their surroundings, aside from the grain-grain contacts, allowing for clean, high-precision access to the basic physics of granular contact charging. This is particularly important in cases where the charging rate is so small that  subtle differences in the initial condition of the grains (material, hydrophilicity, surface chemistry) have a significant effect on the overall charging behaviour. 

The acoustic-levitation-based technique we demonstrate here is material-independent, and can be extended straightforwardly to other particle types, surface chemistries, shapes, and sizes. Since the dynamics of the particle-particle collision are controlled by the acoustic trap, our setup could also be extended to probe the effect of collisional velocity and spin on collisional charging.  Our method to trigger controlled collisions between levitated objects is highly general, and serves as a platform for further studies of non-equilibrium assembly, as well as applications in containerless processing. 

\begin{acknowledgments}
We thank Victor Lee for his contributions to the early stages of this project, Endao Han for suggesting the use of the LED to measure the instantaneous electric field, and Grayson Jackson for his assistance with the particle preparation. We thank Kieran Murphy, Leah Roth, and Adam Wang for useful discussions. This research was supported by the National Science Foundation through Grant No. DMR-1810390. We acknowledge the Chicago MRSEC for the use of the high-speed cameras through its shared experimental facilities. 
\end{acknowledgments}

\setcounter{section}{0}
\renewcommand{\thesection}{APPENDIX \Alph{section}}

\section{Equilibrium point calculation}
\label{sec:absoluteH}
\setcounter{equation}{0}
 \renewcommand{\theequation}{\Alph{section}\arabic{equation}}
 
 When we recover position data from video frames, the absolute height of the particle from the reflector plate cannot be measured simply. The only reliable measurement we can make is relative positions between frames. This is only an issue because of the nonlinear dependence of the acoustic force \eqref{eq:Fay} on position, shifting $Y_j \to Y_j + \Delta y$ changes $X_{1j}$, but none of the other $\vec{X}_i$. To estimate the correct shift $\Delta y$ to apply to our data, we first approximate the equilibrium position assuming a linearized acoustic potential, then perform a brute-force search in the neighborhood of this guess using the full nonlinear potential. The optimal equilibrium position then uniquely determines the necessary shift $\Delta y$. In properly shifted coordinates, the true equilibrium position is given by
\begin{align}\label{eq:eq_pos}
    y_0(a) = \frac{\pi}{2k}-\frac{1}{2k} \arcsin\lrp{\frac{g}{a}}\,.
\end{align}

Thus, in principle, shifting $Y_j \to Y_j - \avg{Y}_t + y_0(a)$ solves this issue, but doing so causes the equation of motion to become nonlinear in $a = \theta_1$ making standard linear regression inappropriate. Note that $\avg{\cdot}_t$ denotes averaging over the time index. To begin, we obtain an estimate for $a$ from the resonance frequency of the particle. Following Lee et al.\cite{lee2018collisional} the instantaneous frequency $f(t)$ and amplitude $A(t)$ of the particle trajectory are found via Hilbert transform. Taking the resonance time to be $\argmax_t A(t) = t_{res}$, then approximating the acoustic force as linear, we find
\begin{align*}
    a \approx \frac{(2\pi f(t_{res}))^2}{2k}\,.
\end{align*}

To correct this initial guess, we perform linear regression on the trajectory for a set of fixed values $\varphi \equiv (1/2k) \arcsin(g/a)$ in a small neighborhood. We then take $\varphi_0$, the value of $\varphi$ leading to the best regression, as our estimate. Quantitatively, we define $X_1(\vec{Y},t_j)$ as Eq. \eqref{eq:Fay} evaluated on the shifted position data:
\begin{align*}
    X_1(\vec{Y},t_j) = \sin(2k(Y_j - \avg{Y}_t) - \varphi_0)\,.
\end{align*}

When fitting on simulated data, the time domain used in regressions for this preliminary minimization is irrelevant. In real data however, different time domains seem to yield different estimates for $\varphi_0$. We elected to always fit over a 2s window immediately preceding $t_{res}$ for a few reasons. First, as we will discuss in Appendix C, velocity and acceleration data need to be re-scaled to reverse errors induced by finite difference effects. This re-scaling approximation is frequency dependent and is most dependable when the amplitude is growing due to driving near resonance. Second, within this regime Fig. \ref{fig:charge_fitting} illustrates that the scale of acceleration due to acoustic force is several decades above all other effects.
 
\section{Electric field fitting}
\setcounter{equation}{0}
 \renewcommand{\theequation}{\Alph{section}\arabic{equation}} 
 In order to couple $q$ to acceleration in our equation of motion, we need to know the electric field in the cavity as a function of time. As shown in Fig. \ref{fig:charges}(a,c), the electric field strength is encoded frame-by-frame as the reflection of an LED on the transducer surface which is visible in the upper portion of the picture. When the LED is on, the $\vec{E}$ field has a positive vertical component which is approximately proportional to its brightness. The challenge we are presented with is fitting a model of the electric field to this data. We know that the electric field is a swept sine with a total sweep time $T$, initial frequency $f_i$, and final frequency $f_f$. The instantaneous frequency is
\begin{align*}
    \omega_E(t) = 2\pi \lrp{\frac{\alpha}{2}t + f_i}t, \quad \alpha = \frac{f_f - f_i}{T}\,.
\end{align*}

This model admits a prediction for the time points $\tau_n$ corresponding to the peaks of the electric field, which are the most easily extracted feature of the data. Because $\vec{X}_4$ is the only term coupling the charge $q$ to the data, this fit needs to be extremely precise, and two seemingly small effects must be taken into account. The first is a slight phase shift in the signal which accounts for the finite frame rate: since the function generator is not synced with the camera, the sweep trigger almost always falls at some point in time between two exposures. The second effect is a minuscule error in the scale of time as a result of very slightly different clock speeds between the camera and function generator. We therefore introduce two fit parameters, a phase shift $\psi$ and time scale $t_s$, which we will vary over in an attempt to minimize the sum of squared errors in observed ($\tilde{\tau}_n$) and calculated ($\tau_n$) peak times:
\begin{equation}
    \begin{aligned}
    \tau_n(\psi,t_s) &= -\frac{f_i}{t_s\alpha} + \frac{1}{t_s} \sqrt{\lrp{\frac{f_i}{\alpha}}^2 + \frac{1}{\alpha}\lrp{\frac{\psi}{\pi} + \frac{1}{2} + 2 n}}\\
    (\psi, t_s) &= \argmin_{\psi,\ t_s}\sum_n (\tilde{\tau}_n - \tau_n(\psi,t_s))^2
    \end{aligned}
\end{equation}

A typical value for this time scaling in our setup is around  $t_s\approx 0.99996$. After carrying out this minimization, we can evaluate $X_4(t)$.
\begin{align}
    X_4(t) = \frac{E_0}{m}\sin(\omega_E(t_st)t_st - \psi)
\end{align}

\section{Finite difference rescaling}
\setcounter{equation}{0}
 \renewcommand{\theequation}{\Alph{section}\arabic{equation}}
 
 Throughout the sweep, each particle's trajectory is approximately harmonic with frequency not too different from the frame rate: $1/f \sim 10\Delta t $. Because these rates are similar, finite difference estimates for velocity and acceleration underestimate the true values. We correct for this by re-scaling the velocity and acceleration according to the error induced at maximal values: 
\begin{align*}
    \dot{y} \to g_1 \dot{y},\quad \ddot{y} \to g_2 \ddot{y}\,.
\end{align*}

For a central first-order difference acting on a sinusoid, the error appears at the zeros of the signal:
\begin{align*}
    g_1(\omega) = \frac{\omega \Delta t}{\sin \omega \Delta t} \, .
\end{align*}

The second order difference introduces error at the peaks of the signal:
\begin{align*}
    g_2(\omega) = \frac{\omega^2 \Delta t^2}{2 - 2\cos \omega \Delta t} \, .
\end{align*}

This rescaling is frequency dependent, and we approximate the response frequency as the driving frequency $\omega_E(t)$. If $\Delta^{(1)}$ and $\Delta^{(2)}$ are our first- and second-order central finite difference operators respectively, then we can define:
\begin{align*}
    V_j(\vec{Y}) &= g_1(\omega_E(t_j)) \lrs{\Delta^{(1)}\vec{Y}}_j\\
    X_2(\vec{Y},t_j) &= 2V_j\\
    X_3(\vec{Y},t_j) &= 2|V_j|V_j
\end{align*}

Similarly,
\begin{align*}
    Z_j(\vec{Y}) &= g_2(\omega_E(t_j)) \lrs{\Delta^{(2)}\vec{Y}}_j + g
\end{align*}
As we take $\omega\Delta t\to 0 $, we find $g_1 \to 1$ and $g_2 \to 1$. If it can be achieved, this is a preferable condition, as this re-scaling is only an approximation and varies in validity throughout the course of a sweep. This highlights an important consideration for future iterations of this experiment. For experiments with stronger acoustic potentials, a proportionally higher frame rate will be necessary to keep this error in check. 
 
\section{Error in measured charging rate} 
\setcounter{equation}{0}
 \renewcommand{\theequation}{\Alph{section}\arabic{equation}}
 
 Equation~\eqref{eq:regression_err} includes an approximation for $\Sigma_s$, the variance contributed by uncertainty in the pixel scale measurement. This does not affect each $\theta_i$ equally: due to approximate linearity of $X_1$ and $X_2$ on $\vec{Y}$, does not significantly impact $\theta_1$ or $\theta_2$. When $s$ is the pixel scale and $\sigma_s$ is its uncertainty, we have
\begin{align}
    \Sigma_s = \left(\frac{\sigma_s}{s}\right)^2
    \begin{bmatrix} 
    0 & ~ & ~ & ~\\
    ~ & 0 & ~ & ~\\
    ~ & ~ & \theta_3^2 & \theta_3\theta_4  \\
    ~ & ~ & \theta_4\theta_3 & \theta_4^2
    \end{bmatrix}\,.
\end{align}

To express the variances for charging rate fit parameters, we first need to define more terms. Let the charging data be $\vec{q} = (q_1,\dots,q_{N_q})^T$ with $\delta \vec{q}$ defined similarly. Let $C$ be a 2\by $N_q$ matrix such that $C_{1j}$ is the number of collisions preceding the measurement of $q_j$ and $C_{2j} = 1$. Then for our charging rate fit parameters $\vec{\gamma}$, we have
\begin{align*}
    \vec{q} = C^T \vec{\gamma} + W^{-1}\vec{\eta}\,.
\end{align*}
with $W_{ii} = 1/\delta q_i$. Thus
\begin{equation}\label{eq:regression_gamma}
    \begin{aligned}
        \vec{\gamma} &= (CW^2C^T)^{-1}CW^2\vec{q}\\
        \Sigma_\gamma &= (CW^2C^T)^{-1}\lrp{CW + \frac{\eta^2}{N_q - 2}}\,.
    \end{aligned}
\end{equation}

We then calculate the error in charging rate using the diagonals of $\Sigma_\beta$ to find 
\begin{align*}
    \delta \gamma_i = \sqrt{(\Sigma_{\gamma})_{ii}}\,.
\end{align*}

\section{Estimate of the maximum contact area during a collision}
\setcounter{equation}{0}
 \renewcommand{\theequation}{\Alph{section}\arabic{equation}}
We estimate the maximum contact area~$A_c$ during a head-on collision between two elastic spheres of radii $R_1$ and $R_2$ using Hertzian contact theory. 


Upon collision, the potential energy stored in elastic deformation will be\cite{landau1960theory,timoshenko1970theory}
\begin{align}
    U = \frac{2}{5}\varepsilon r^{1/2}h^{5/2}\,.
\end{align}

Where $(2R_1 + 2R_2) - h$ is the total distance between the centers of mass of the particles at maximal compression, and the reduced radius $r$ and elastic constant $\varepsilon$ are given by
\begin{equation}
    \begin{aligned}
    r &= \frac{R_1R_2}{R_1 + R_2}\\
    \epsilon &= \frac{4}{3}\lrp{\frac{1-\sigma_1^2}{E_1} + \frac{1-\sigma_2^2}{E_2}}^{-1}\,.
    \end{aligned}
\end{equation}

In this expression, $E_i$ are the Young's moduli and $\sigma_i$ are the Poisson coefficients. From momentum conservation,
\begin{align}
    h = h_1 + h_2 = h_1\lrp{1 + \frac{m_1}{m_2}}\,,
\end{align}

where $h_1$ is the depression into particle 1, and $h_2$ is defined similarly. Substituting this expression for $h$ yields
\begin{align}
    h_1 = \lrp{1 + \frac{m_1}{m_2}}^{-1}\lrp{\frac{5U}{2\varepsilon}r^{-1/2}}^{2/5}\,.
\end{align}

In the center of mass frame, we can simply substitute the total kinetic energy in for~$U$. Finally, working within the approximation of a small deformation, 
\begin{align}
    A_1 \approx 2\pi R_1 h_1\,.
\end{align}

In the special case that both particles have the same material parameters and radius~$R$, we find
\begin{align}
    A = 2\pi\lrs{\frac{15}{16}\frac{mv^2(1-\sigma^2)}{E}R^2}^{2/5}\,.
\end{align}

Using videos of collisions to estimate the kinetic energies of particles prior to the collision and equating this to the total elastic potential energy, we find that $A \sim 1000 \mu\mathrm{m}^2$.
%

\end{document}